\documentclass[twocolumn, superscriptaddress, prl]{revtex4-1}
\usepackage{graphicx}
\usepackage[integrals]{wasysym} 
\usepackage{nicefrac,amssymb,amsmath} 

\usepackage{color}
\newcommand{\GF}{Green's function}
\newcommand{\QP} {quasiparticle}

\newcommand{\beq}{\begin{equation}}
\newcommand{\eeq}{\end{equation}}

\begin{document}

\title{Dispersing and non-dispersing satellites in the  photoemission spectra of aluminum   }

\newcommand{\lsi}{Laboratoire des Solides Irradi\'es, \'Ecole Polytechnique, CNRS, CEA,  Universit\'e Paris-Saclay, F-91128 Palaiseau, France}
\newcommand{\etsf}{European Theoretical Spectroscopy Facility (ETSF)}
\newcommand{\soleil}{Synchrotron SOLEIL, L'Orme des Merisiers, Saint-Aubin, BP 48, F-91192 Gif-sur-Yvette, France}
\newcommand{\pmc}{Laboratoire de Physique de la Mati\`ere Condens\'ee, \'CNRS Ecole Polytechnique,  Universit\'e Paris-Saclay, F-91128 Palaiseau, France}
\newcommand{\seattle}{Department of Physics, University of Washington, Seattle, Washington 98195-1560, USA}
\newcommand{\sls}{Swiss Light Source, Paul Scherrer Institut, CH-5232 Villigen PSI, Switzerland}
\newcommand{\insp}{Sorbonne Universit\`e, CNRS, Institut des Nanosciences de Paris, UMR7588, F-75252, Paris, France}

\author{Jianqiang Sky Zhou}
\affiliation{\lsi}
\affiliation{\insp}
\affiliation{\etsf}

\author{Lucia Reining}
\affiliation{\lsi}
\affiliation{\etsf}

\author{Alessandro Nicolaou}
\affiliation{\soleil}

\author{Azzedine Bendounan}
\affiliation{\soleil}

\author{Kari Ruotsalainen}
\affiliation{\soleil}

\author{Marco Vanzini}
\affiliation{\lsi}
\affiliation{\etsf}

\author{J. J. Kas}
\affiliation{\seattle}

\author{J. J. Rehr}
\affiliation{\seattle}

\author{Matthias Muntwiler}
\affiliation{\sls}

\author{Vladimir N. Strocov}
\affiliation{\sls}

\author{Fausto Sirotti}
\affiliation{\pmc}

\author{Matteo Gatti}
\affiliation{\lsi}
\affiliation{\etsf}
\affiliation{\soleil}

\begin{abstract}
Satellites in electronic spectra are pure many-body effects, and their study has been of increasing interest in both experiment and theory. The presence of
satellites due to plasmon excitations can be understood with simple models of electron-boson coupling.  It is far from obvious how to match such a model to real spectra, where more than one kind of quasi-particle and of satellite excitation coexist. Our joint experimental and theoretical study
shows that satellites in the angle-resolved photoemission spectra of the prototype simple metal aluminum consist of a superposition of dispersing and non-dispersing
features. Both are due to electron-electron interaction, but the non-dispersing satellites also reflect the thermal motion of the atoms. Moreover,
besides their energy dispersion, we also show and explain a strong shape dispersion of
the satellites. By taking into account these effects,
our first principles calculations using the GW+C approach of many-body perturbation theory reproduce and explain the experimental spectra 
to an unprecedented extent.
\end{abstract}

\date{\today}%
\maketitle

Photoemission spectroscopy is one of the most direct experimental tools to access  band structures and excitation spectra of materials \cite{Huefner2003}. Although the Coulomb interaction between electrons leads to a renormalization of energies and to lifetime broadening, the resulting \QP\ (QP) band structure can usually still be detected, and, moreover, reproduced and interpreted using first-principles theoretical approaches \cite{Martin2016,Onida2002,Aryasetiawan1998,qpscgw-prl2006}. However, the QPs constitute only  part of the measured spectra. They are usually accompanied by an incoherent background and a series of additional structures called satellites, over a binding-energy range of several tens of eV. These structures are pure many-body effects \cite{Martin2016}. They cannot, by definition, be interpreted from a single-particle point of view. Therefore, they carry information that is complementary to the insight gained from the band structure. 

The simplest picture for the origin of photoemission satellites is given by an electron-boson coupling model, which describes the excitation of one or more bosons when an electron is removed from the sample  \cite{Langreth1970,Chang1972,Hedin1999}.
The effective shake-up bosons may represent e.g. plasmons, magnons or phonons, which are distinguished by their characteristic energy and dispersion. The model yields a Poisson series of equidistant satellites separated by the boson energy. When the bosonic excitation is dispersing, this distance is a weighted average of its energies \cite{Hedin1971,Hedin1999}.

In the framework of first-principles calculations, the physics of this model is captured by the GW+C approach \cite{Aryasetiawan-cumulant1996}, and in its analogue for phonons \cite{Verdi2017}. This approach has recently gained considerable renewed popularity \cite{cumulant-Ferdi-Al-experiment1999,Aryeasetiawan-al-cumulant2002,Vos2001,Kheifets2003,Guzzo-prl2011,Guzzo2012, Steven-cumulant2013,Gatti2013,Guzzo2014,Josh-RC2014,Fabio-cumulant-2015,zhou-jcp2015,Lischner2015,Mayers2016,Caruso2016,Gumhalter2016,Nakamura2016,ISI:000423115200010,Borgatti2018,Seidu2018,Caruso2018}. In the GW approximation \cite{Hedin-GW1965}, the electronic QP is screened by charge excitations contained in the dynamically screened Coulomb interaction $W$. These excitations correspond to the bosons, and the Poisson series is created by the cumulant (C) expansion, where the GW self-energy appears as the main ingredient in the exponent of an exponential representation of the \GF , with peaks roughly corresponding to those in the loss function $\sim |{\rm Im}  W|$. Similarly, the role of a boson can be played by a phonon propagator (see e.g. \cite{Mahan1966,Dunn1975,Skinner1982,Hsu1984,Mahanbook,*[{}] [{ and references therein.}] Nery2018}). The method has been used to explain a variety of experimentally observed plasmonic and phononic satellites from first principles. 
Satellite energies are overall well reproduced. Moreover, angle-resolved studies show that the plasmon satellite dispersion roughly
 follows that of the valence bands \cite{Kheifets2003,Fabio-cumulant-2015,Lischner2015,Kas2014}, as one would expect when considering excitations at each point in the Brillouin zone to be approximately independent \cite{Guzzo-prl2011}.  Individual angle-resolved spectra of real materials, however, contain much more information than just dispersing bands and replica due to one kind of boson. In the present joint experimental and theoretical work we show that to describe and analyze such spectra requires major steps forward. 

We have performed state-of-the-art angle-resolved photoemission spectroscopy (ARPES) measurements as a function of photon energy and first-principles GW+C calculations of the prototype simple metal  aluminum \cite{Levinson1983,Ma1994,Krasovskii2008}, yielding the band structure and the first three dispersing satellites. Whereas the qualitative picture of dispersing satellites is confirmed, our work also demonstrates that the standard GW+C calculations alone cannot explain single angle-resolved spectra. Instead, we show in which way, besides the intrinsic angle-resolved spectral function, additional contributions due to thermal motion of the atoms and to
inelastic scattering of the photoelectrons strongly contribute to the measured spectra. Our results confirm the quantitative success of the GW+C approach for the intrinsic spectral function, allow us to highlight and explain the shape dispersion of the satellites, and elucidate the origin of strong non-dispersing satellites.

ARPES experiments  were performed at about 50 K 
using a Scienta EW4000 hemispherical electron analyzer at the PEARL beamline 
of the Swiss Light Source \cite{ISI:000391724900038}. The Al($\bar{1}01$) single crystal sample surface was prepared with several cycles of 
sputtering and annealing until the observation of a carbon 
contamination below 1\% of a monolayer and a sharp low-energy electron diffraction pattern.
Photon-energy dependent ARPES experiments were used to measure the Fermi surface periodicity in the $k_z$ direction and to identify the photon energies  corresponding to the $\Gamma$ points of 4 consecutive Brillouin zones (see supplemental material \cite{suppmat}). ARPES experiments on a 60 eV binding energy range were then performed at 293 eV, 624 eV and 1100 eV photon energies, corresponding to small $k_z$ deviations from the ideal $\Gamma$L direction \cite{suppmat}. 
Intrinsic spectral functions along $\Gamma$L (with $ 0.2$ $(0.9)$ eV shifts which account for deviations in the $k_z$ direction for 624 (1100) eV ARPES) are calculated from first principles using the GW+C approach described in \cite{zhou-jcp2015,ISI:000423115200010}. 
Computational details can be found in  \cite{suppmat}. 

\begin{figure}[tb]
 \includegraphics[width=\columnwidth,keepaspectratio=true]{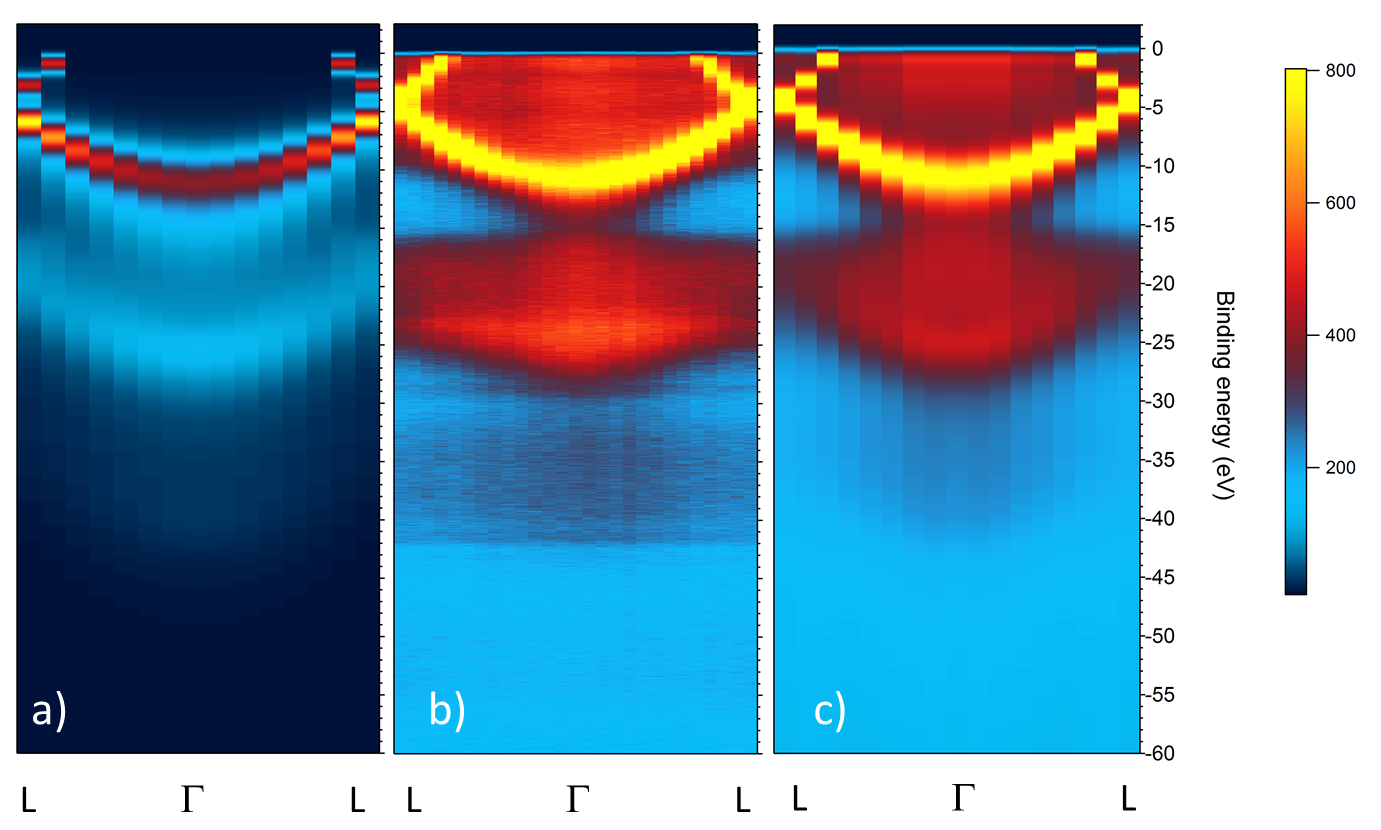}
 \caption{(Left and right panels) Calculated spectral functions (shifted along $k_z$ as detailed in \cite{suppmat}) and (middle panel) experimental ARPES image in the $\Gamma$L direction.
The results of \textit{ab initio} GW+C calculations are without  (left panel) and with (right panel) the Debye-Waller contribution and extrinsic and interference effects. }
 \label{fig:overview}
\end{figure}

The ARPES image obtained at 624 eV photon energy is shown in the middle panel of Fig. \ref{fig:overview}. One can clearly see the parabolic valence band of aluminum and its replica at a distance of about 15 eV. This is consistent with the bulk plasmon energy of aluminum \cite{Kloos1973}.
The GW+C theoretical spectra, shown in the left panel of Fig. \ref{fig:overview}, are in qualitative agreement with experiment concerning the valence bands and the presence of a dispersing satellite, similar to previous observations for silicon  \cite{Kheifets2003,Fabio-cumulant-2015,Lischner2015}. Moreover, the existence of other satellite replicas is confirmed, as expected from GW+C. Still, the comparison of theoretical and experimental panels shows large differences. In order to say more, one has to compare single angle-resolved spectra, which is much more demanding than a comparison of intensity plots. An attempt for Al was made in \cite{cumulant-Ferdi-Al-experiment1999,Aryeasetiawan-al-cumulant2002}, by comparing results of GW+C calculations and electron momentum spectroscopy, but this could only probe spherically averaged spectra. In comparison to GW alone, it was shown that GW+C is in much better agreement with experiment. The remaining significant discrepancies still left a full explanation of the angle-resolved spectra as an open problem.

 \begin{figure}[tb]
\includegraphics[width=\columnwidth,keepaspectratio=true]{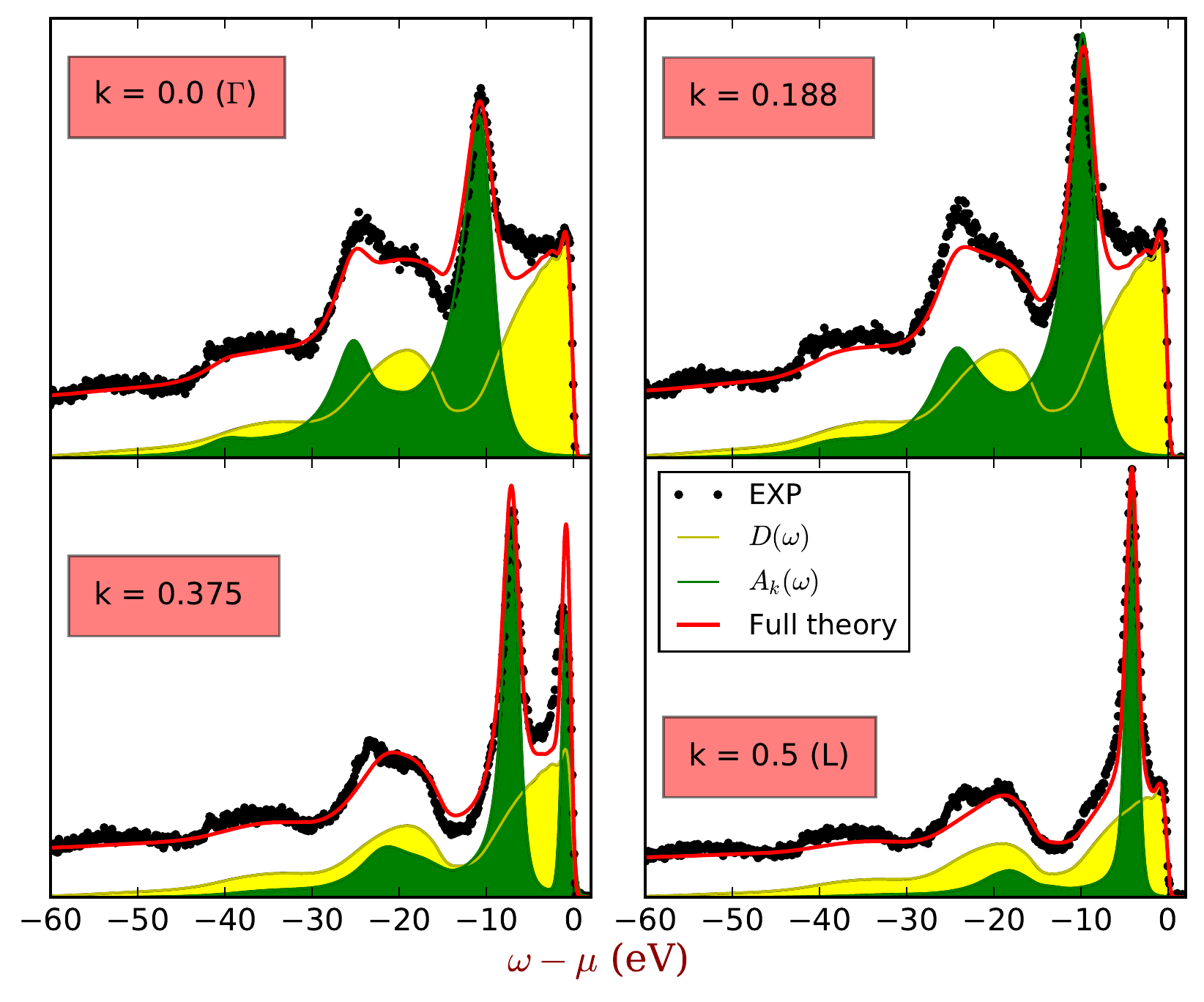}
 \caption{The spectra at 624 eV photon energy for four k points along $\Gamma$L. $\mu$ is the Fermi energy. The sum of the intrinsic angle-resolved spectral function $A_k(\omega)$ (green area) and the angle-integrated $D(\omega)$ (yellow area)
weighted by the Debye-Waller coefficients and with the Shirley background yields the theoretical spectrum (red line), which is compared with ARPES data (black dots).
 }
 \label{fig:SPF-6243d}
\end{figure}

To face this challenge, Fig. \ref{fig:SPF-6243d} shows the same theoretical intrinsic spectral functions (green areas) as the left panel of Fig. \ref{fig:overview}, and the measured spectra (black dots) at 624 eV photon energy for several ${\bf k}$ points along $\Gamma$L. However, in these fully angle-resolved results,  the agreement is extremely disappointing, both concerning the valence-band region and the satellites. While some of the experimental structures can be identified in the calculated spectra, in particular the QPs at $\Gamma$ and L, and part of the satellite structures, others are completely absent in the calculation: for example, the shoulder that is measured on the first satellite around -20 eV.
Since there are no absolute intensities in the experiment, the agreement might  be  improved by scaling the intensity or by taking into account a secondary electron background in the calculated curves, but the qualitative disagreement remains. This is in contrast with the conclusions of numerous publications using GW+C for angle-integrated spectra \cite{Guzzo-prl2011,Steven-cumulant2013}  or angle-resolved intensity plots such as \cite{Fabio-cumulant-2015,Lischner2015,Verdi2017}, where the agreement was sufficiently good to answer the questions posed by those studies. Instead, the large differences in individual spectra in Fig. \ref{fig:SPF-6243d} hinder any further discussion and analysis.

We attribute the source of discrepancy to the presence of a strong non-dispersing, but energy-dependent, background. In the band-structure region,
such a background has been discussed  as a consequence of Debye-Waller (DW) effects \cite{Shevchik1977,Shevchik1979,Hussain1980, PhysRevB.88.205409,Sondergaard2001,Hofmann2002}. Even at temperatures as low as 50 K in aluminum, the thermally disordered motion of the atoms weakens the momentum conservation of direct photoelectron transitions and hence partly destroys the momentum resolution of the ARPES. The effect increases with increasing temperature and photon energy, thus being one of the limiting factors for the angle resolution in hard x-ray photoemission spectroscopy experiments \cite{ISI:000329376900007,Gray2011,Plucinski2008,Papp2011}.
Since aluminum is a metal, the loss of angular resolution explains the fact that even in our experimental spectrum at $\Gamma$, which corresponds to the bottom of the band, much intensity is found at the Fermi level. Other possible explanations such as surface photoelectric effects \cite{Plummer1987,Hansen1997} and the $k_z$ intrinsic broadening of the final state \cite{Strocov2003} were discarded here since insufficient  to reproduce this additional intensity.

While the importance of DW on the band-structure region is established, the effect on
satellites, to the  best of our knowledge, has heretofore not been investigated.
The question is therefore whether, and if yes in which way, the DW physics is also responsible for strong modifications of the satellites. It is reasonable to surmise that plasmon satellites are created in the same way in ordered and thermally disordered systems. 
This would mean that the angle-integrated QP component due to the DW effect leads to angle-integrated satellites, in the same way as for the angle-resolved QPs and their satellites.

In order to test this  idea, we have to determine the magnitude of the angle-integrated spectral component. As a further complication, in the intrinsic spectral functions (green areas in Fig. \ref{fig:SPF-6243d}) extrinsic and interference effects due to inelastic scattering of the outgoing photoelectron are not included \cite{Mahan1973}. The magnitude and consequences of these effects have been subject to a vivid discussion \cite{Pardee1975,Attekum1978,Huefner2003}. Recently the model approach of Refs. \cite{Hedin1999,Bardyszewski1985,Hedin1998} was combined with first-principles calculations of spectral functions to discuss  angle-integrated spectra of silicon \cite{Guzzo-prl2011,Guzzo2012}, and comparison with experiment was significantly improved with respect to a purely intrinsic calculation. One may expect that the inelastic scattering of the outgoing electrons from the angle-resolved QPs also 
leads to a loss of angular resolution \cite{Bardyszewski1985}. Therefore, we add the extrinsic and interference contributions to the angle-integrated spectral function due to the DW effect, similar to the approaches used in  \cite{Guzzo-prl2011,Guzzo2012} for angle-integrated spectra \cite{suppmat}.
In order to determine the relative weight of angle-resolved and angle-integrated spectra, 
we use the fact that only the angle-integrated part can contribute weight at the Fermi level for the ${\bf k}$ points shown here. The ratio of angle-resolved versus angle-integrated contributions in the theoretical spectra  used here is about 1:1.5, which is consistent with estimates from DW theory \cite{Hussain1980,suppmat}.

\begin{figure}[tb]
\includegraphics[width=\columnwidth,keepaspectratio=true]{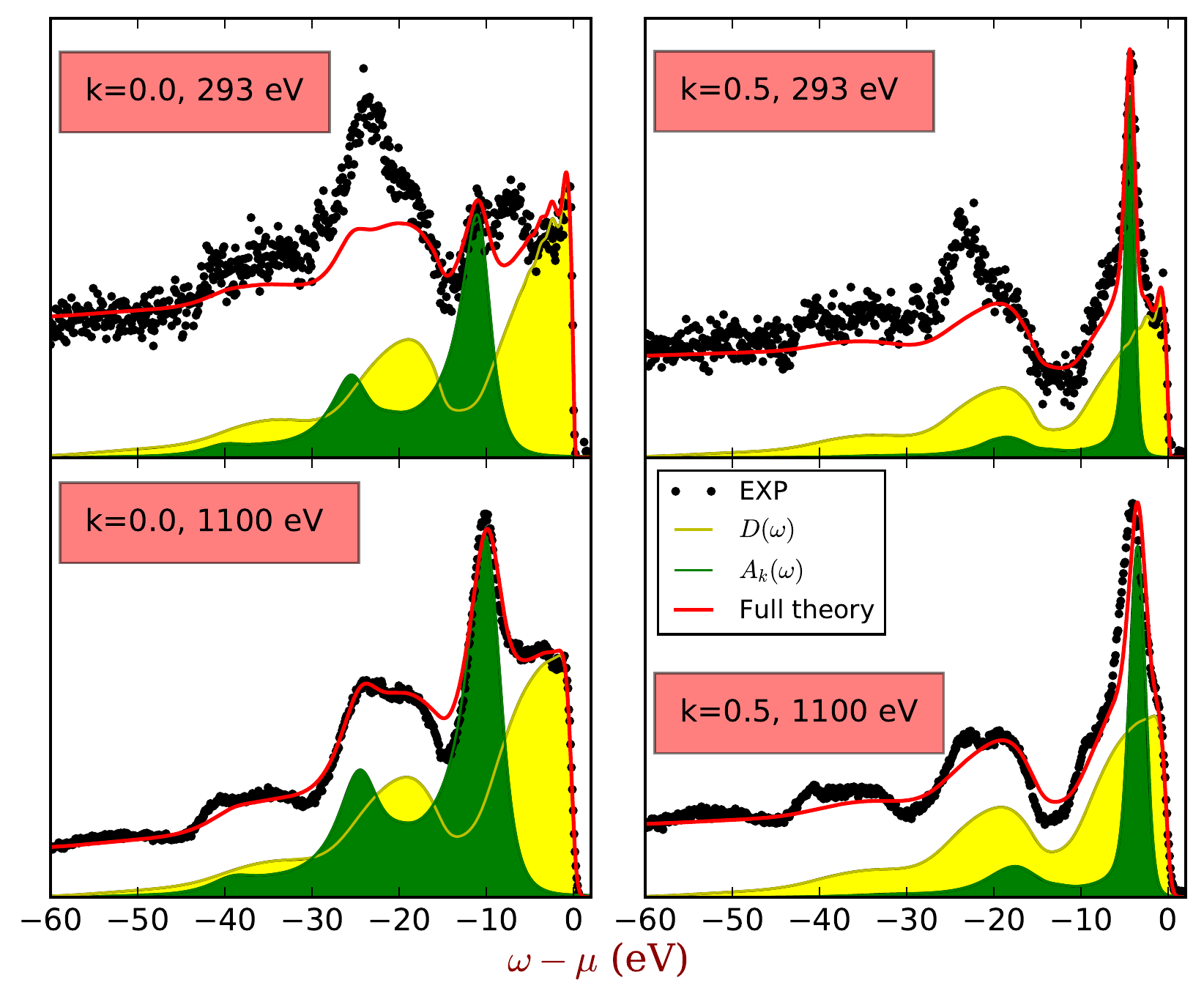}
 \caption{The spectra at 293 eV and 1100 eV photon energy for the $\Gamma$ and L points. The theoretical spectra (red lines) are obtained as in Fig. \ref{fig:SPF-6243d}.}
 \label{fig:SPF-293-1100-2d}
\end{figure}

The red curves in Fig. \ref{fig:SPF-6243d}  are our final result, including cross sections, secondary electron background and Fermi functions \cite{suppmat}. 
The agreement with ARPES data is excellent. Our conjecture is further confirmed by the  trends expected for the DW effect when changing photon energy: 
Fig. \ref{fig:SPF-293-1100-2d} shows results for photon energies of 293 eV and 1100 eV.  For the latter, the angle-integrated contribution increases to yield a ratio between 1:2 and 1:3, which is  in agreement with DW estimates. For the former, the surface dominates. The effective Debye temperature is smaller at the surface than in the bulk \cite{Sondergaard2001,Hofmann2002,Petersen1999}. This compensates the photon energy dependence, leading to a ratio of 1:3, again as expected \cite{Sondergaard2001,Hofmann2002}.

The enhanced surface sensitivity might also explain a pronounced structure in the QP spectrum around $-8$ eV absent from the calculations, which could be due to a surface contribution, and which might be related to the sharp structure on the first satellite. Indeed, it is much less visible at higher photon energies.
Detailed numbers can be found in \cite{suppmat}.

Our calculation of extrinsic and interference effects is based on an approximate model \cite{Hedin1999,Bardyszewski1985,Hedin1998}, and one of the aims of the present work is to get more insight about its validity. In this model, which also includes surface plasmons, the imaginary part of the GW self-energy at a frequency shifted by the QP energy, ${\rm Im} \Sigma_{xc}(\omega+\varepsilon)$, determines the magnitude of the cumulant. It is multiplied by a frequency ($\omega$)- and photon energy ($\nu$)-dependent enhancement factor $R_{\nu}(\omega)$ derived from the electron gas. This factor expresses extrinsic and interference effects: for $R_{\nu}=1$ only the intrinsic spectral function contributes. We have found that the model $R_{\nu}(\omega)$ accurately describes the spectra at 624 eV, but for 293 and 1100 eV better results are obtained by introducing a scaling factor $\mathcal{C}$ such that $R_{\nu}(\omega) \to \mathcal{C}(R_\nu(\omega)-R_{\nu}(0))+R_{\nu}(0)$ where $\mathcal{C}=1.2$ and $0.7$ are adopted for 293 eV and 1100 eV, respectively. Still, these moderate deviations from the ideal model value of $\mathcal{C}=1$, and the fact that $\mathcal{C}$ is the only ajustable parameter used here, indicate that effects which are neglected in the current approach, such as recoil effects in GW+C \cite{Hedin1980}, are either small or approximately canceling. 

\begin{figure}[tb]
 \includegraphics[width=0.8\columnwidth,keepaspectratio=true]{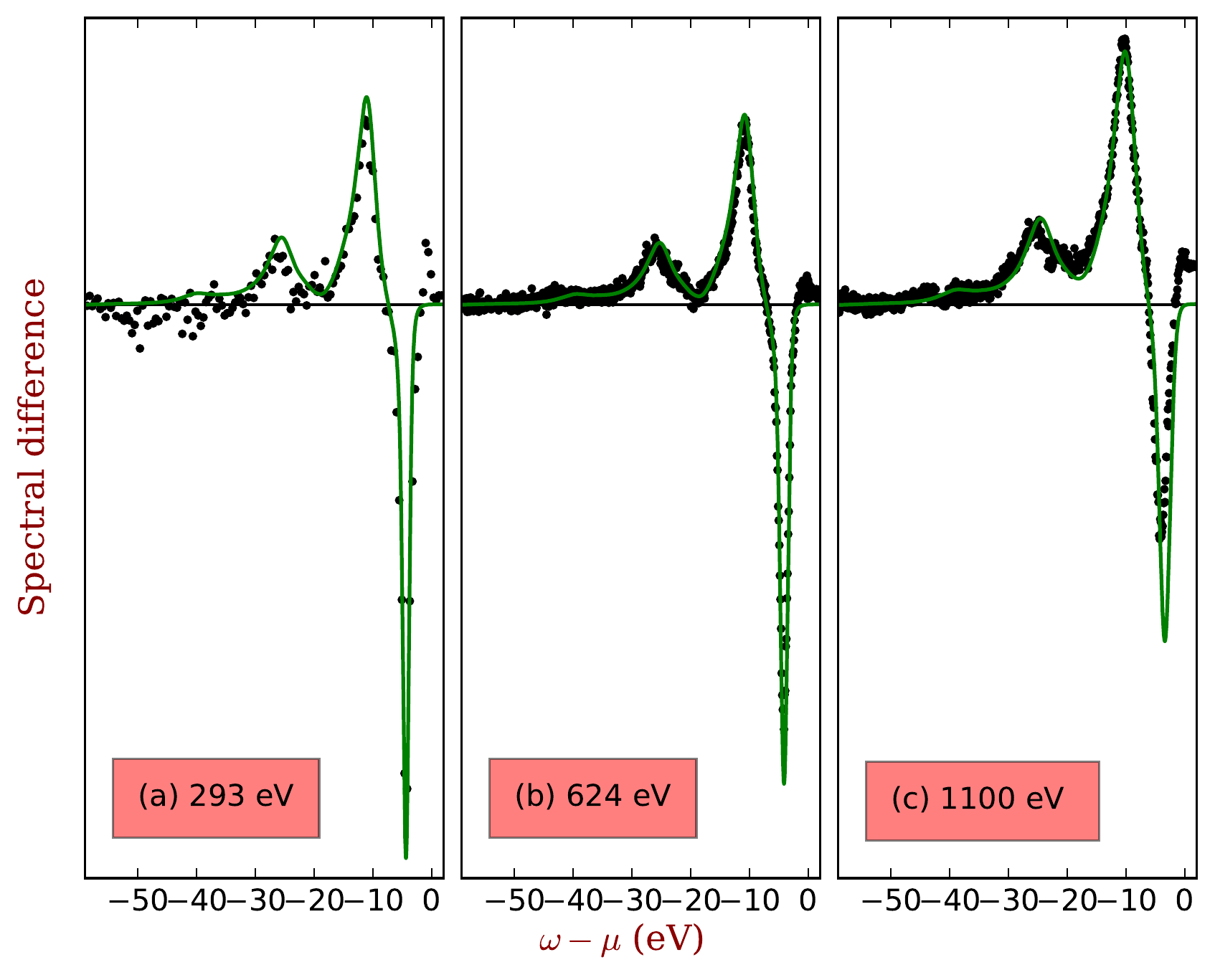}
 \caption{Difference between spectra at the $\Gamma$ and L points, at 293,  624 and 1100 eV photon energy: experimental results (black dots) are compared with calculations (green lines). The horizontal black lines indicate the offsets due to the secondary electron background (see text).}
 \label{fig:TritticoDiff}
\end{figure}

The excellent agreement between theory and experiment, as reflected in the right panel of Fig. \ref{fig:overview}, is quite satisfactory, but Fig. \ref{fig:SPF-6243d} finally shows how difficult it is to make a comparison between intrinsic spectral functions and angle-resolved experiments. In the following we propose a way to overcome this difficulty.
One first has to distinguish between the non-dispersing and the dispersing features in the experiment. To this end, we take the difference between spectra measured at different angles and we consider the most different spectral functions, which are observed at $\Gamma$ and L. In these two directions the {angle-resolved} photoemission intensity at the Fermi level is zero. 
We then normalize the two experimental spectra  such that their \textit{non-dispersing} intensity at the Fermi level is the same (293 eV data were binned over 9 points to reduce the noise). The resulting difference spectrum has an almost constant offset starting from the QP region, which is related to the secondary electron background. Therefore we apply a constant vertical shift in order to compare the experimental difference spectra to the calculated ones, which are normalized to match the experimental QP intensities. The result is shown in Fig. \ref{fig:TritticoDiff} for the three photon energies.

The negative feature close to the Fermi energy is due to the QP peak at L, and the following positive one is the QP at $\Gamma$. A clear satellite difference structure is found between $-15$ eV and $-30$ eV.  All features are very well represented by the calculated difference of intrinsic spectral functions. This implies that the calculations yield correctly not only the position, but also the shape dispersion of the satellites. This shape dispersion is indeed quite strong, as one can see clearly from green areas in Fig. \ref{fig:SPF-6243d}. 
The first intrinsic satellite is strong and sharp at $\Gamma$, whereas it is much more washed out at L. 
This can be understood by looking at the GW self-energy matrix elements for an occupied state $\ell$:
\begin{equation}
 {\rm Im} \Sigma_{xc}^{\ell\ell}(\omega) = \sum_{j,s\neq0}|V_{{\ell}j}^s|^2 \delta(\omega - \varepsilon_j + \omega_s)
 \, ,
 \label{eq:model-sigma-Hedin}
\end{equation}
where $|V_{{\ell}j}^s|$ are matrix elements of the 
fluctuation potentials \cite{Hedin1999}, and $\omega < \mu$. Peaks of $ {\rm Im} \Sigma(\omega)$ determine the satellites. Energy conservation allows contributions at a QP energy $\varepsilon_j$ minus a bosonic excitation  $\omega_s$ (e.g. a plasmon).
Because of $|V_{{\ell}j}^s|$, the dominant contributions in the sum are given by QP states with states 
$j$ similar to $\ell$ and by plasmon excitations at small momentum $\mathbf{q}$. This gives rise to very different satellite shapes as a function of $\mathbf{k}$. In particular, at  L  a long tail of one-electron states available at energies $\varepsilon_j < \varepsilon_{\ell} $ induces a  satellite broadening towards larger binding energies that is absent at the bottom of the band at $\Gamma$. In other words, the characteristic shape evolution of the satellites reflects the density of valence states, and this behavior is very well captured by the GW+C approach.

In conclusion, our joint experimental and theoretical study 
allows us to analyze the photoemission spectra of  aluminum well beyond the simple picture of bands and their plasmon replicas. First, we have found a shape dispersion of the ARPES satellites, which we could discuss by taking difference spectra, and which is explained by the density of valence states. Second, the presence of dispersionless satellites is due to the thermal motion of the atoms, which induces a loss of momentum conservation for both the QPs and their satellites. These dispersionless satellites can be  sharp in the valence spectra, corresponding to replicas of  van Hove singularities in the density of states, and they should not be confused with signals due to impurities. Third, inelastic scattering of the photoelectron due to the electron-electron interaction enhances the angle-integrated satellites, in agreement with model predictions. The photon-energy dependence of the measured and calculated spectra confirms our conjecture. The insight gained from this work clarifies the role of various physical processes, and suggests new opportunities to extract qualitative and quantitative information about materials from ARPES.

Computation time was granted by GENCI (Project No. 544). This work has received funding from the European  Research Council (ERC Grant Agreement n. 320971) and from a Marie Curie FP7 Integration Grant within the 7th European Union Framework Programme. This work was also supported by a public grant overseen by the French National Research Agency (ANR) as part of the ``Investissements d'Avenir'' program (Labex NanoSaclay, reference: ANR-10-LABX-0035). We acknowledge the Paul Scherrer Institut, Villigen, Switzerland for synchrotron radiation beamtime at the PEARL beamline of the Swiss Light Source.

\bibliographystyle{apsrev4-1} 
\bibliography{alu-bib}
\end{document}


\title{Supplemental Material for: \\ Dispersing and non-dispersing satellites in the  photoemission spectra of aluminum }

\newcommand{\lsi}{Laboratoire des Solides Irradi\'es, \'Ecole Polytechnique, CNRS, CEA,  Universit\'e Paris-Saclay, F-91128 Palaiseau, France}
\newcommand{\etsf}{European Theoretical Spectroscopy Facility (ETSF)}
\newcommand{\soleil}{Synchrotron SOLEIL, L'Orme des Merisiers, Saint-Aubin, BP 48, F-91192 Gif-sur-Yvette, France}
\newcommand{\pmc}{Laboratoire de Physique de la Mati\`ere Condens\'ee, \'CNRS Ecole Polytechnique,  Universit\'e Paris-Saclay, F-91128 Palaiseau, France}
\newcommand{\seattle}{Department of Physics, University of Washington, Seattle, Washington 98195-1560, USA}
\newcommand{\sls}{Swiss Light Source, Paul Scherrer Institut, CH-5232 Villigen PSI, Switzerland}
\newcommand{\insp}{Sorbonne Universit\`e, CNRS, Institut des Nanosciences de Paris, UMR7588, F-75252, Paris, France}

\author{Jianqiang Sky Zhou}
\affiliation{\lsi}
\affiliation{\insp}
\affiliation{\etsf}

\author{Lucia Reining}
\affiliation{\lsi}
\affiliation{\etsf}

\author{Alessandro Nicolaou}
\affiliation{\soleil}

\author{Azzedine Bendounan}
\affiliation{\soleil}

\author{Kari Ruotsalainen}
\affiliation{\soleil}

\author{Marco Vanzini}
\affiliation{\lsi}
\affiliation{\etsf}

\author{J. J. Kas}
\affiliation{\seattle}

\author{J. J. Rehr}
\affiliation{\seattle}

\author{Matthias Muntwiler}
\affiliation{\sls}

\author{Vladimir N. Strocov}
\affiliation{\sls}

\author{Fausto Sirotti}
\affiliation{\pmc}

\author{Matteo Gatti}
\affiliation{\lsi}
\affiliation{\etsf}
\affiliation{\soleil}


\maketitle

\section{Experimental details}
\label{sec:exp}

The Al single crystal was prepared with several sputtering and annealing cycles to obtain a clean surface (C contamination less than 1\%).
The sample contained  about 0.02$\%$ atom concentration of tantalum impurities from the sample holder. Because of their extremely high cross section, Ta 4$f$ photoelectrons  show up as a sharp doublet line in the spectrum at the characteristic binding energies of  21.12 and  20.25 eV. Two Gaussian peaks at these  energies  have thus been removed in the angle-resolved spectra.

\begin{figure}[hbt]
 \includegraphics[width=0.9\columnwidth,keepaspectratio=true]{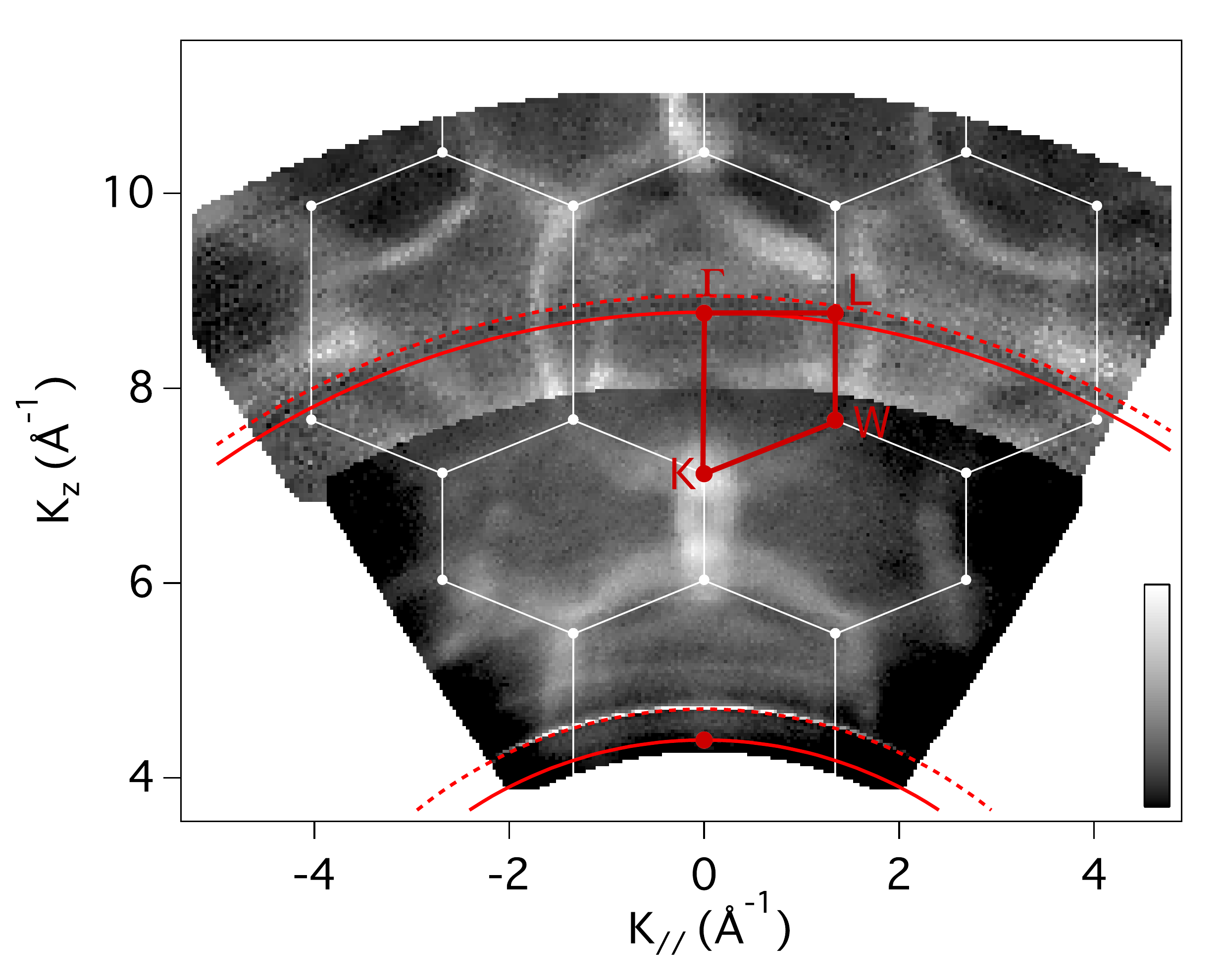}
 \caption {Cut of the Fermi surface in the $\Gamma$LWK plane
 obtained by scanning the photon energy, allowing the identification of the Brillouin zones (white lines) and the high-symmetry points (thick red line indicating a $\Gamma$LWK path). The ARPES experiments at the two lowest photon energies have been performed along the two red dashed lines, which introduces small deviations from the ideal $\Gamma$L path (see text).} 
 \label{fig:fermi}
\end{figure}

In order to deduce the periodicity of the electronic structure in the $\Gamma$LWK plane, ARPES experiments were performed in the range of photon energies between 60 and 350 eV. They were used to build the Fermi surface presented in Fig. \ref{fig:fermi}. An inner potential $V_0=17$ eV, with a work function $\Phi=4.5$ eV, was determined in order to match the position of the Fermi surface with the Brillouin zone contours. 
Four consecutive $\Gamma$ points were identified\cite{Himpsel1980} corresponding to the following photon energies: 61 eV, 281 eV, 649 eV, 1163 eV.
The photon energies used to obtain the ARPES spectra over a binding energy of 60 eV, namely 293 eV, 624 eV and 1100 eV (see main text), correspond to small shifts along the surface-normal direction $k_z$, which at $\Gamma$ were evaluated respectively as: -0.1, 0.15 and 0.29 of $\Gamma$K (and similarly for the other points along $\Gamma$L). 

\section{Calculated intrinsic spectral functions}

For an introduction to the theoretical framework we refer e.g. to Ref. \onlinecite{Martin2016}. Here we provide the computational details for the calculation of the intrinsic spectral functions.

\subsection{Computational details}
\label{sec:intrinsic}

Calculations within density-functional theory in the local-density approximation (LDA) and the GW approximation of many-body perturbation theory  have been carried out using the ABINIT package \cite{abinit-code}. Cumulant expansion calculations were done with the Cumupy code \cite{cumupy,zhou-prb2018}.
The experimental crystal structure\cite{Wyckoff1963} was adopted.
In order to take into account the polarization from $2s$ and $2p$ semicore states\cite{zhou-prb2018},  a norm-conserving Troullier-Martins pseudopotential \cite{pseudop-PRB1993} was built that treats $2s$ and $2p$ semicore states explicitly as valence electrons. The plane-wave cutoff was 260 Hartree and the Brillouin zone (BZ) was sampled using a $16\times16\times 16$ $\Gamma$-centered $\mathbf{k}$-point grid. 
A smearing of 0.005 Hartree was used in order to speed up the $\mathbf{k}$-point convergence. 

The self-energy within the partially energy-self-consistent GW approximation (updating the energies in the Green's function $G$, but keeping the screened Coulomb potential $W$ fixed) was obtained starting from the LDA. 
$W$ was calculated in the adiabatic local-density approximation (ALDA) with 80 bands, 2000 and 200 plane waves for the wavefunctions and dielectric matrix, respectively,  
on a mesh consisting of 120 frequencies up to 45 eV along the real axis and 10 frequencies along the imaginary axis.
The intraband transitions in the dielectric function for $\vb{q}=0$ were taken into account  using\cite{Cazzaniga2012,Cazzaniga-PRB2010-intraband} $\epsilon_{intra}(\omega) = 1 - \omega_p^2/[\omega(\omega+\imi \eta)]$, where the parameters $\omega_p=15.03$ eV and $\eta=0.58$ eV were fitted\cite{zhou-prb2018} on the calculated retarded loss function for small $\vb{q} \neq 0$.
In the calculation of the self-energy,  80 bands, and 2000 plane-waves for both wavefunctions and the exchange term were used. 
10 bands were updated in the energy self-consistent cycle.
The GW QP band structure is in agreement with previous results from Refs. \cite{Northrup1989,Bruneval2006,Cazzaniga2012}.

With the GW self-energy matrixelements $\Sigma_{xc}^{\ell\ell}(\omega)$, the time-ordered cumulant expansion \cite{zhou-jcp2015,zhou-prb2018} of diagonal elements of the one-particle Green's function $G_\ell$ for each occupied state $\ell$ ($\mu$ is the Fermi energy) was obtained from:
\begin{equation}\label{Eq:C-toc-h}
\begin{split}
 G_\ell(\tau) & = \mathrm{i}\theta(-\tau) e^{-\mathrm{i}\varepsilon_\ell\tau}e^{C_\ell(\tau)}\\ 
  C_\ell(\tau) & = \frac{1}{\pi} \int_{-\infty}^{\mu-\varepsilon_\ell} \dd\omega \abs{{\rm Im} \Sigma_{xc}^{\ell\ell}(\omega+\varepsilon_\ell)} \frac{e^{-\mathrm{i}\omega \tau} - 1}{\omega^2} \, ,
 \end{split}
\end{equation}
where the QP energy $\varepsilon_\ell < \mu$ was obtained in the GW calculation.
The imaginary part of the Fourier transform of $G_\ell$ in the frequency domain yields the intrinsic spectral function $A_\ell(\omega)=\pi^{-1}\abs{\Im G_\ell(\omega)}$. 

We have taken into account the experimental deviation from the ideal $\Gamma$L line (see Sec. \ref{sec:exp})
by calculating the corresponding energy shifts obtained from the band dispersion along $k_z$. For the three photon energies we have used the following  values: 0 eV for 293 eV photon energy, 0.26 eV for 624 eV, and 0.9 eV for 1100 eV.

\section{Full photoemission spectra}

In the following we explain how to simulate the measured photoemission spectra starting from the calculation of the intrinsic spectral function (see Sec. \ref{sec:intrinsic}). 

\subsection{Cross sections and normalisation}

The photoemission experimental spectra are in arbitrary units with respect to theoretical spectra. In other words, one always has to set an overall factor, which does not influence any discussion of \textit{differences} of spectra. Ideally there would be no other parameter. However, in practice both state-of-the-art theory and experiment introduce an uncertainty. 
On the experimental side, the measured intensity shows a $\mathbf{k}$-dependent (but energy independent) variation 
which can be due, for example, to photoelectron diffraction effects \cite{Vicente1996} or to the detector efficiency change as a function of the slit angle.
On the theory side, photoemission cross sections have to be taken into account.
For angle-integrated spectra they can be usually estimated combining tabulated values of atomic photoionization cross sections (see e.g. \cite{Yeh1985}) and calculated projected densities of states. This procedure is less reliable for valence bands of delocalized-electron materials like aluminum, especially when different bands are compared through one or more Brillouin zones, like in the present case, where two bands are occupied along the $\Gamma$L direction.

\begin{figure}[tb]
\includegraphics[width=1.3\columnwidth,keepaspectratio=true]{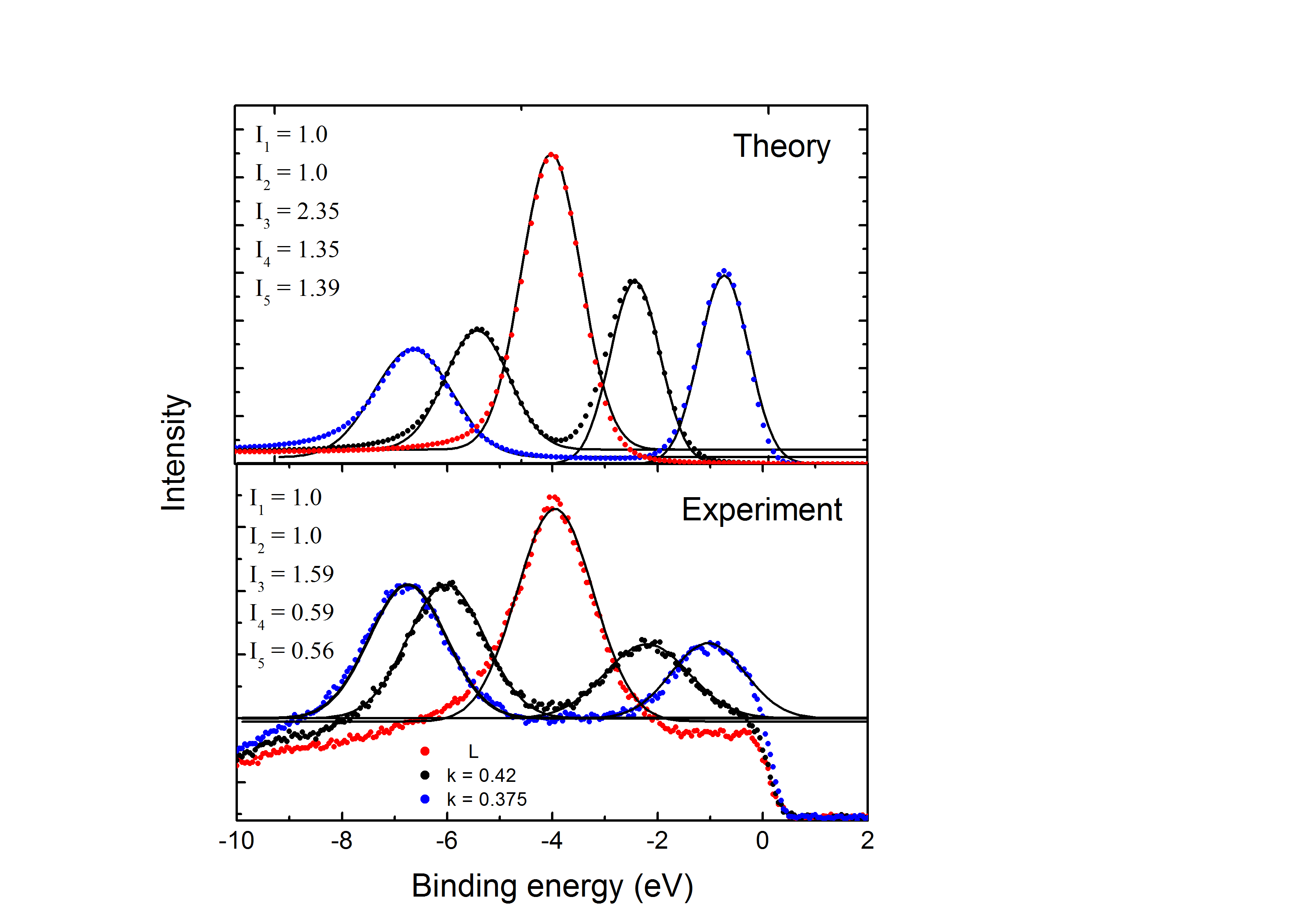}
 \caption{Angle-resolved spectra calculated (top) and measured (bottom) for three different $k$ values along the $\Gamma$L direction. The theoretical and measured lineshapes (data points) are reproduced with Gaussian functions (continuous lines). In the spectra at $k=0.42$ and $k=0.275$ (black and blue curves, respectively) the two peaks correspond to two different bands. Their intensity ratios have been fixed in the calculations in order to match the corresponding experimental ratios (see text).}
 \label{fig:crosssections}
\end{figure}

Both problems can be overcome by using solely the QP part of the spectra, as follows. 
First we chose a ${\bf k}$-point between $\Gamma$ and L where two bands are occupied. By requiring the intensity ratio of the two bands to be the same as in experiment (see Fig. \ref{fig:crosssections}), starting from the tabulated cross sections \cite{Yeh1985}, we determined a correction factor of 0.4 for the theoretical cross section of the second band. As we have verified, this factor does not change for other ${\bf k}$-points along $\Gamma$L, which supports our hypothesis.

\subsection{Extrinsic and interference effects}

\begin{figure}[hbt]
\includegraphics[width=0.7\columnwidth,keepaspectratio=true]{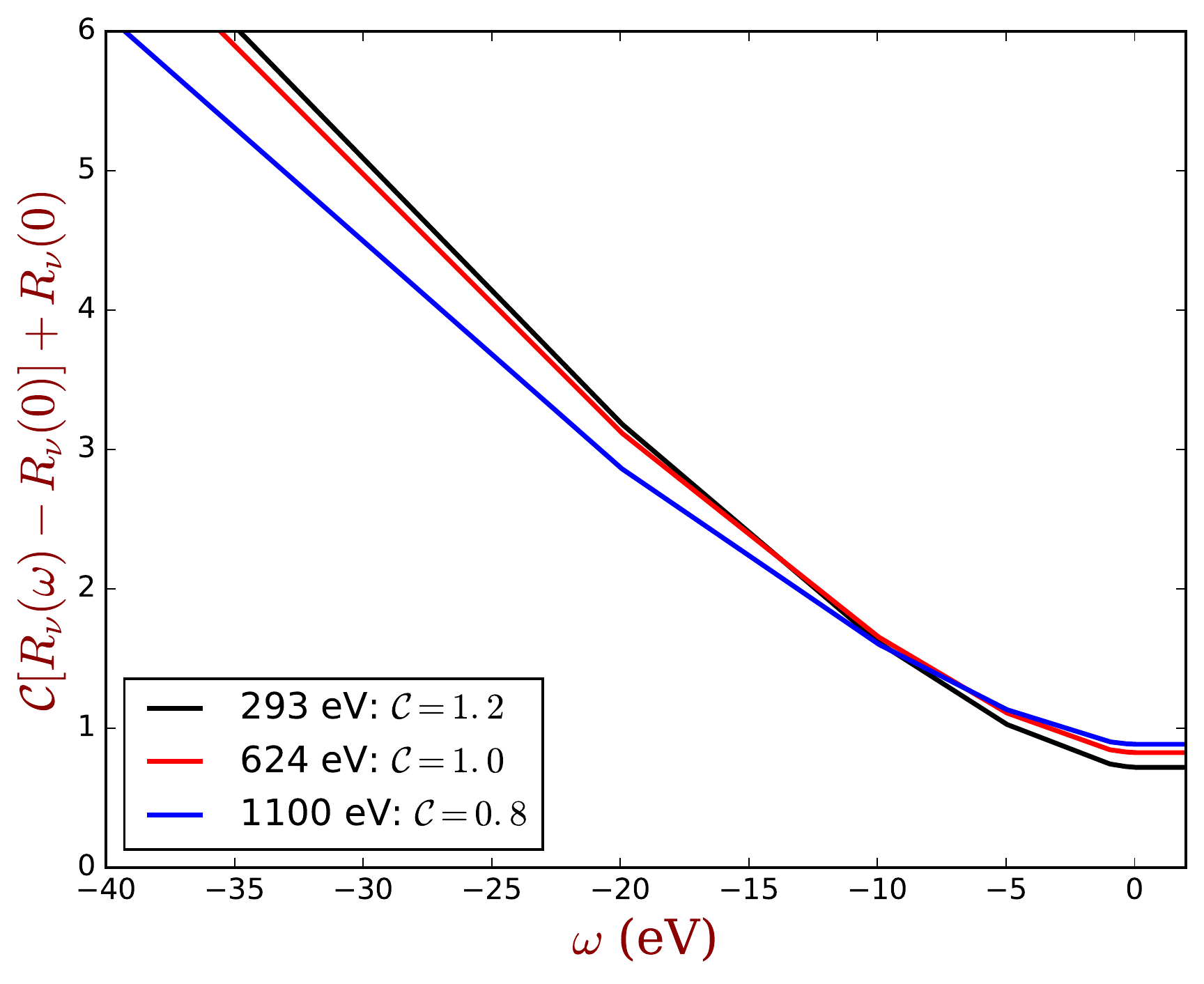}
 \caption{The extrinsic and interference function $\mathcal{C}(R_\nu(\omega)-R_{\nu}(0))+R_{\nu}(0)$ used for the three photon energies. }
 \label{fig:conclusion-extrinsic}
\end{figure}

The photoelectron on the way out from the sample to the detector can induce other bosonic excitations (such as plasmons). This process is called extrinsic effect to distinguish it from the intrinsic excitation by the photohole.  A further contribution, of opposite sign, is due to the quantum interference between extrinsic and intrinsic processes.
Both extrinsic and interference effects are photon-energy dependent and they are not contained in the intrinsic spectral function.
A way to approximately include them in the calculation of photoemission spectra was recently followed in Refs. \onlinecite{Guzzo-prl2011,Guzzo2012,zhou-jcp2015}
by combining first-principles cumulant calculations with a model developed by Hedin and coworkers \cite{Bardyszewski1985,Hedin1998}.
The resulting angle-integrated photoemission spectra of silicon and sodium were largely improved with respect to purely intrinsic calculations\cite{Guzzo-prl2011,zhou-jcp2015}.
These model calculations provide the function $R_{\nu}(\omega)$, displayed in Fig. \ref{fig:conclusion-extrinsic}, representing the sum of  extrinsic and interference effects for different photon energies $h\nu$.
The spectra including extrinsic and interference effects are obtained by  multiplying ${\rm Im} \Sigma_{xc}^{\ell\ell}(\omega+\varepsilon_\ell) $ in Eq. \eqref{Eq:C-toc-h} by $R_{\nu}(\omega)$.
We have found that the calculated function $R_{\nu}(\omega)$ had to be rescaled by a factor $\mathcal{C}$ depending on the photon energy (see Tab. \ref{tab:photon-energy-gb} and Fig. \ref{fig:conclusion-extrinsic}), to be precise: $\mathcal{C}(R_\nu(\omega)-R_{\nu}(0))+R_{\nu}(0)$.
The spectra including extrinsic and interference effects have been used to obtain the integrated density of states that corresponds to the non-dispersing contribution to the final spectra (see Sec. \ref{sec:DW}). 
We refer to Sec. \ref{sec:extrinsic} for further analysis.

\subsection{Debye-Waller effect}
\label{sec:DW}

The disordered thermal vibrations of the atoms destroy the $\mathbf{k}$-selectivity of the ARPES, giving rise to a non-dispersive component in the spectrum which amounts to integrating the different contributions from the $\mathbf{k}$-resolved spectra  over the BZ (Debye-Waller effect).
In the simplest approximation, the ${\bf k}$-resolved  and ${\bf k}$-integrated components are weighted by the Debye-Waller factors $W$ and $1-W$, respectively. 
The Debye-Waller factors can be approximately estimated\cite{Hussain1980} as: $W = \exp[-\frac{1}{3}{\Delta}k^2{\langle}U^2(T)\rangle]$,
where ${\Delta}k$ is the momentum transfer at the photon energy of the measurement and  ${\langle}U^2(T)\rangle$ is the three-dimensional mean-squared vibrational displacement that can be estimated knowing the Debye temperature of the material and the temperature $T$ of the sample.  According to the model, for increasing photon energy and temperature, the ${\bf k}$-integrated spectrum tends to become the dominant component.

In the following, for each state $\ell=n{\bf k}$, we call $A_\ell$ the calculated ${\bf k}$-resolved spectrum and $D(\omega)$ the calculated $\mathbf{k}$-integrated spectrum.
We call $\alpha_\ell$ and $\beta_\ell$, respectively, the weights that we have to give to the  ${\bf k}$-resolved  and ${\bf k}$-integrated components in order to match the measured spectra.
These weights have been  determined by the condition that at the Fermi level only the ${\bf k}$-integrated component contributes.
The resulting numbers are gathered in Tabs. \ref{tab:624-3d}-\ref{tab:293-1100-3d}.
 Since the experimental spectra are in arbitrary units, the $\alpha_\ell$ and $\beta_\ell$ values contain an arbitrary scaling factor, and only their ratio is important. This ratio should be directly linked to the Debye-Waller factor $W$ by $W=(\alpha_\ell/\beta_\ell)/(1+\alpha_\ell/\beta_\ell)$. Indeed, the values of $W$ that we obtain from our fit, namely 0.4 and 0.25 for 624 and 1100 eV photon energy, respectively, are consistent with the corresponding estimates of $W=0.53$ and $W=0.33$, respectively, from the model \cite{Hussain1980}. In particular, the decrease with increasing photon energy is similar. The comparison for the lowest photon energy, 293 eV, where the surface becomes relevant, is not possible, since the simple model Ref. \onlinecite{Hussain1980} does not take into account the surface. Indeed, since the effective Debye temperature at the surface is smaller than in the bulk, the fit to the experiment yields a strong effect, namely $W=0.26$, whereas the model would predict $W=0.75$ for Al at T=77 K, corresponding to a pure bulk contribution.

Finally, the ${\bf k}$-resolved and the $\mathbf{k}$-integrated components,
$A_\ell(\omega)$ and $D(\omega)$, are also supplemented by the corresponding Shirley backgrounds\cite{Shirley1972}  of secondary electrons, $B_\ell(\omega)$ and $B_D(\omega)$.  

For each state $\ell$ the final calculated spectrum $S_\ell(\omega)$ thus reads:
\begin{multline}
S_\ell(\omega) = f(\omega) \{ \alpha_\ell[A_\ell(\omega) + c_\ell B_\ell(\omega)] \\ 
+ \beta_\ell [D(\omega) + c_\ell B_D(\omega)] \}\, ,
\end{multline}
where the coefficient $c_\ell$ is fixed by the condition that the calculated spectrum matches the tail of the measured spectrum at about -60 eV (where $A_\ell(\omega)$ and $D(\omega)$ are vanishing) and  the Fermi function 
\begin{equation}
f(\omega) =\left[1+ e^{\frac{\omega-\mu}{\sqrt{(kT)^2 + K^2}}}\right]^{-1}
\end{equation}
is evaluated at $T=50$ K and contains an additional parameter $K$ 
that accounts for the broadening introduced by the energy resolution of the experimental setup. 
The angle-resolved $A_\ell(\omega)$ and the angle-integrated $D(\omega)$ have both been broadened with a Gaussian function.
The widths used for the various photon energies are reported in Tab. \ref{tab:photon-energy-gb}.

\begin{table}[ht!]
\caption{\label{tab:624-3d}%
Coefficient of Shirley background $c_\ell$ and ratio of Debye-Waller factors $\alpha_\ell/\beta_\ell$ at 624 eV}
\begin{ruledtabular}
\begin{tabular}{ l c c}
 state &$c_\ell$ &$ \alpha_\ell/\beta_\ell $ \\
\colrule
$k=0$     & $0.01$ & $0.67$  \\ 
$k=0.188$ & $0.011$ & $0.67$  \\
$k=0.375$ & $0.012$ & $0.67$  \\
$k=0.5$   & $0.014$ & $0.67$ \\ 
\end{tabular}
\end{ruledtabular}
\end{table}

\begin{table}[ht!]
\caption{\label{tab:293-1100-3d}%
Coefficient of Shirley background $c_\ell$ and ratio of Debye-Waller factors $\alpha_\ell/\beta_\ell$ at 293 eV and 1100 eV}
\begin{ruledtabular}
\begin{tabular}{ l c c}
 state &$c_\ell$ &$ \alpha_\ell/\beta_\ell $  \\
\colrule
$k=0, 293 $eV & $0.027$ & 0.357  \\ 
$k=0.5, 293$ eV  & $0.035$ & 0.33  \\
$k=0, 1100$ eV & $0.008$ & $0.52$   \\
$k=0.5, 1100$ eV   & $0.019$ & 0.35  \\ 
\end{tabular}
\end{ruledtabular}
\end{table}

\begin{table}[ht!]
\caption{\label{tab:photon-energy-gb}%
Broadening parameters and rescaling factor $\mathcal{C}$ for extrinsic/interference effects}
\begin{ruledtabular}
\begin{tabular}{ l c c c c}
 photon energy (eV) & Gaussian (eV) & $K$ (eV) &$\mathcal{C}$\\
\colrule
$293$   & $0.4$  & $0.1$ &1.2\\ 
$624$  & $0.5$ & $0.2$ &1.0\\
$1100$  & $0.8$ & $0.3$ &0.8  
\end{tabular}
\end{ruledtabular}
\end{table}

\section{Further analysis of the spectra}
In the following, we give more details about how to obtain reliable information from the comparison of theory and experiment.

\subsection{Intrinsic versus interference/extrinsic contributions}
\label{sec:extrinsic}

\begin{figure}
 \begin{center}
\includegraphics[width=\linewidth]{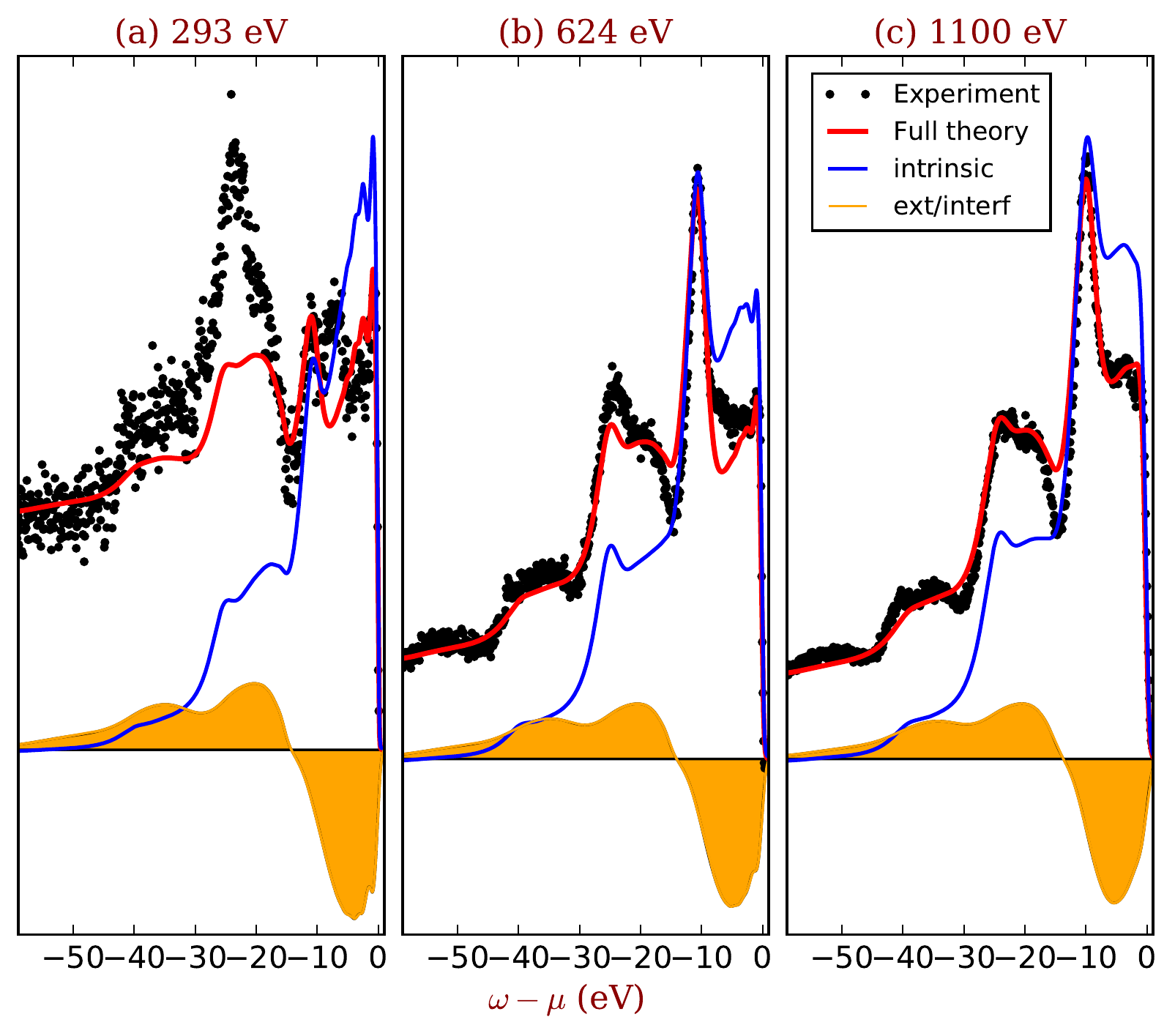}
    \caption{Black dots: experimental spectra at the $\Gamma$ point. Blue line: intrinsic spectral function (with dispersing and non-dispersing components). The calculated and experimental intensities are scaled to match the intensity of the QP peak. Orange area: extrinsic plus interference contribution. Red line: total calculated spectrum (including the background). }
    \label{fig:ext-int}
  \end{center}
\end{figure}

The model we have used to estimate extrinsic and interference contributions to the spectra is quite rough. Nevertheless, our results show that, once the DW contribution is taken into account, the model can explain the difference between calculated intrinsic spectral functions and the experiments to a large extent. In order to give a better idea of the extrinsic and interference contributions, 
Fig. \ref{fig:ext-int} shows the comparison between experiment and theoretical results, which are decomposed into the intrinsic spectral function on one side (blue line), and the extrinsic and interference contributions on the other side (yellow shaded area). Note that the total theoretical result also contains the Shirley background. 

Due to the inelastic scattering of the outgoing electron, the number of electrons that arrive at the detector with higher kinetic energy, and that are therefore interpreted in the spectrum as electrons with lower binding energy, decreases. Instead, these electrons transfer energy to other excitations (here in particular plasmons), and therefore arrive at the detector with less kinetic energy. In this way they contribute to the spectral weight counted at higher binding energy. Indeed, the orange shaded area in Fig. \ref{fig:ext-int} is negative in the QP region and positive in the satellite region. Without this contribution, the agreement between theory and experiment would be quite poor. Note that the model itself is parameter free. With respect to the model, we introduce one single multiplicative parameter $\mathcal{C}$ which, as shown in table \ref{tab:photon-energy-gb}, is always close to unity, which means that the model alone already describes the extrinsic and interference effects almost quantitatively. The figure also shows that the photon energy  dependence is quite mild over the range considered here.

\subsection{Shape of the satellites}

We can understand the shape of the satellites and their change as a function of $\mathbf{k}$
by analysing the matrix elements of the lesser GW self-energy entering the cumulant expansion Eq. \eqref{Eq:C-toc-h}:
\begin{equation}
 {\rm Im} \Sigma_{xc}^{\ell\ell}(\omega + \varepsilon_\ell) = \sum_{j,s\neq0}|V_{{\ell}j}^s|^2 \delta(\omega  + \varepsilon_\ell - \varepsilon_j + \omega_s ) \, .
 \label{eq:model-sigma-Hedin}
\end{equation}
Here $\varepsilon_j$ are single-particle GW energies, $\omega_s$ the boson energies (corresponding to neutral charge excitations such as plasmons and interband transitions) and $|V_{{\ell}j}^s|$ are the fluctuation potentials, determining the strength of the electron-boson couplings.

\begin{figure}
 \begin{center}
\includegraphics[width=0.9\linewidth]{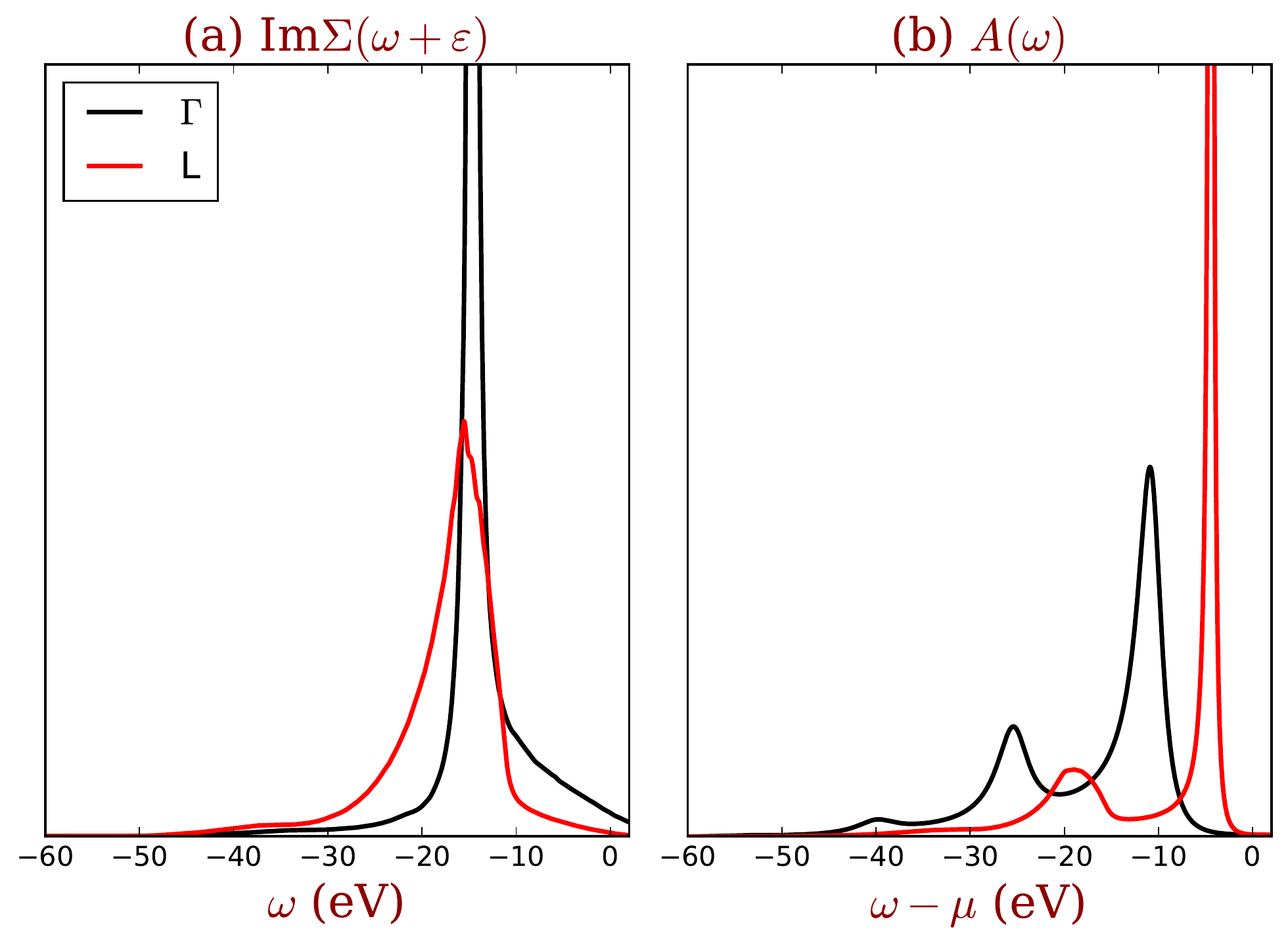}
    \caption{(a) ${\rm Im} \Sigma_{xc}^{\ell\ell}(\omega+\varepsilon_\ell)$  at $\Gamma$ (black) and L (red), (b) The corresponding intrinsic spectral function $A_\ell(\omega)$  at $\Gamma$ and L. }
    \label{fig:sigma}
  \end{center}
\end{figure}

In the simplest case of one electronic level and one boson, the self-energy \eqref{eq:model-sigma-Hedin} has just a delta peak at $\omega_s$ and 
the corresponding cumulant expansion yields a Poisson series of sharp satellite lines with a constant separation equal to the boson energy.
The situation is generally more complicated in a real solid where the electronic and bosonic excitations are dispersing.
In a simple metal like Al, electrons are coupled with the valence plasmon, which has a parabolic dispersion as a function of its wavevector $\mathbf{q}$. In the coupling with a single core level (which has no dispersion),  the plasmon dispersion  produces an asymmetric lineshape in the spectral function with a tail to higher binding energies \cite{Hedin1999}.
The most interesting situation is for the valence band, for which both the single-particle energies $\varepsilon_j$ and the plasmon $\omega_s$ have a non-zero dispersion. 
Fig. \ref{fig:sigma}(a) shows  the calculated ${\rm Im} \Sigma_{xc}^{\ell\ell}(\omega+\varepsilon_\ell)$ for the valence band of aluminum at $\Gamma$ and L. The shape is very different in the two cases: notably the self-energy is much broader at  L than at $\Gamma$. As a consequence, besides the different lifetime broadening of the QP peaks, given by the different values of  ${\rm Im} \Sigma_{xc}^{\ell\ell}(\omega=\varepsilon_{\ell})$, also the satellites in the intrinsic spectral function $A_\ell(\omega)$  have a different shape at $\Gamma$ and $L$: Fig. \ref{fig:sigma}(b) shows that the first satellite at $\Gamma$ is a replica of the QP, whereas the first satellite at L is much broader, and much less peaked, than its QP.
This finding can be explained as follows. Delocalised electronic states as in aluminum are mostly coupled with plasmons with $\mathbf{q}$ close to 0. Indeed, the peaks of ${\rm Im} \Sigma_{xc}^{\ell\ell}(\omega+\varepsilon_\ell)$ in Fig. \ref{fig:sigma}(a) are centred around 15 eV, while the calculated plasmon dispersion in Al ranges from  15  eV at $\mathbf{q}\rightarrow 0$ to  about 25  eV at $\mathbf{q}=1.0$ Bohr$^{-1}$ (see figure 8 in Ref. \cite{Cazzaniga2012-alu-plasmon-disp}). 
Moreover, in the sum \eqref{eq:model-sigma-Hedin} the most important contributions are selected by the electron-boson couplings $V_{{\ell}j}^s$: they are given by the states $j$ that are close to $\ell$.
In the case of $\Gamma$, which is at the bottom of a parabolic valence band, no state can contribute to the sum with energy $\varepsilon_j < \varepsilon_\ell$. As a result, the shape of ${\rm Im} \Sigma_{xc}^{\ell\ell}(\omega+\varepsilon_\ell)$ is asymmetric, with a tail towards smaller binding energies. 
In the case of L, instead, states contributing to the sum can have both larger or smaller energies than $\varepsilon_\ell$ and since the dispersion is much steeper than at the bottom of the band, the shape of ${\rm Im} \Sigma_{xc}^{\ell\ell}(\omega+\varepsilon_\ell)$ becomes at the same time more symmetric and broader. This is then directly reflected in the shape of the satellites in the spectral function.

\bibliographystyle{apsrev4-1} 
\bibliography{alu-supp.bib}

\clearpage


\pagebreak